%% file: main.tex
\documentclass[a4paper, amsfonts, amssymb, amsmath, reprint, showkeys, nofootinbib, twoside, superscriptaddress]{revtex4-2} 
\usepackage[english]{babel}
\usepackage[utf8]{inputenc}
\usepackage[colorinlistoftodos, color=green!40, prependcaption]{todonotes}
\usepackage{enumerate}
\usepackage{lipsum}
\usepackage{braket}
\usepackage{csquotes}
\usepackage{quantikz}
\usepackage[pdftex, pdftitle={Article}, pdfauthor={Author}]{hyperref}
\usepackage[most]{tcolorbox}
\begin{document}
\title{Scalable Quantum State Preparation for Encoding Genomic Data with Matrix Product States}

% Potentially superfluous list of authors - included Lloyd and PIs
% Discuss w/ Sergii who else from project to include

\author{Floyd M. Creevey}
% \email{floyd.creevey@unimelb.edu.au}
\email{fc309@sanger.ac.uk}
\affiliation{School of Physics, University of Melbourne, VIC, Parkville, 3010, Australia.}
\affiliation{Wellcome Trust Sanger Institute, Hinxton, Cambridgeshire, CB10 1SA, United Kingdom.}
% \affiliation{The Department of Applied Mathematics and Theoretical Physics (DAMTP), University of Cambridge, Cambridge, CB2 3EH, United Kingdom.}
\author{Hitham T. Hassan}
\email{hh12@sanger.ac.uk}
\affiliation{Wellcome Trust Sanger Institute, Hinxton, Cambridgeshire, CB10 1SA, United Kingdom.}
\author{James McCafferty}
\email{jdmc@sanger.ac.uk}
\affiliation{Wellcome Trust Sanger Institute, Hinxton, Cambridgeshire, CB10 1SA, United Kingdom.}
\author{Lloyd C. L. Hollenberg}
\email{lloydch@unimelb.edu.au}
\affiliation{School of Physics, University of Melbourne, VIC, Parkville, 3010, Australia.}
\author{Sergii Strelchuk}
\email{sergii.strelchuk@cs.ox.ac.uk}
\affiliation{Department of Computer Science, University of Oxford, Oxford, OX1 3QG, United Kingdom.}
% \author{Richard Durbin}
% \email{rd109@cam.ac.uk}
% \affiliation{Department of Genetics, University of Cambridge, Cambridge, CB2 3EH, United Kingdom.} % likewise would be good to check Richard's 

\date{\today}

\begin{abstract}
    % As quantum computing hardware continues to improve, there is a growing need for algorithms that enable classical data to be loaded into the quantum states of said devices. This work focuses on implementing these theoretical algorithms, specifically for the problem of encoding genomic data into a quantum computer. Recent work using matrix product states (MPS) has provided avenues for this encoding to be done efficiently and effectively. A MPS can be generated that represents arbitrary classical data, which can be used to generate low-depth quantum circuits that encode the data onto quantum devices. Using this implementation a circuit to encode the entire genome of the organism $\Phi X174$ was generated, a world first, as well as much larger genomes.

    As quantum computing hardware advances, the need for algorithms that facilitate the loading of classical data into the quantum states of these devices has become increasingly important. This study presents a method for producing scalable quantum circuits to encode genomic data using the Matrix Product State (MPS) formalism. The method is illustrated by encoding the genome of the bacteriophage $\Phi X174$ into a 15-qubit state, and analysing the trade-offs between MPS bond dimension, reconstruction error, and the resulting circuit complexity. This study proposes methods for optimising encoding circuits with standard benchmark datasets for the emerging field of quantum bioinformatics. The results for circuit generation and simulation on HPC and on current quantum hardware demonstrate the viability and utility of the encoding.

    %marking a world first by generating a quantum circuit for encoding the entire genome of the organism $\Phi X174$, as well as circuits for much larger genomes. 

    % As quantum computing hardware continues to advance, there is an increasing demand for algorithms that facilitate the loading of classical data into the quantum states of these devices. Recent developments utilising matrix product states (MPS) have created effective and efficient methods for this data encoding. A MPS can be generated to represent arbitrary classical data, which can then be used to create low-depth quantum circuits that encode the data onto quantum devices. This work specifically focuses on implementing these theoretical algorithms to encode genomic data into a quantum computer. As a result of this implementation, a circuit was developed to encode the entire genome of the organism $\Phi X174$, marking a world first, as well as accommodating much larger genomes.
\end{abstract}

\keywords{quantum state preparation, quantum computing, bioinformatics, MPS}

\maketitle

\input{sections/introduction.tex}
\input{sections/method.tex}
\input{sections/complexity.tex}
\input{sections/results-hpc.tex}
\input{sections/results-hardware.tex}
\input{sections/conclusions.tex}
\input{sections/acknowledgements.tex}

\bibliography{MPS.bib}

% \onecolumngrid

% \appendix*
% \input{sections/appendix1.tex}

\end{document}

%% file: sections/introduction.tex
\section{Introduction} \label{sec:introduction}

Efficient algorithms to encode genomic data into quantum states open novel avenues to accelerate the complex workflows of computational genomics with quantum computing. For example, the quantum sequence alignment (QSA) algorithm~\cite{hollenberg_fast_2000}, which gains a quadratic speedup over its classical counterpart, requires data to be encoded into a superposition state within the quantum computer before execution of the algorithm. This highlights the importance of efficient and effective encoding of classical data into quantum computers as an essential first step in quantum bioinformatics pipelines. Constructing circuits that prepare states for genomic data will enable fully quantum pipelines for complex bioinformatics tasks to be implemented with quantum algorithms. This includes such algorithms as the previously mentioned QSA algorithm~\cite{hollenberg_fast_2000} for sequence alignment, implemented in Ref.~\cite{Creevey:2025qsr}, other quantum bioinformatics algorithms developed since (see~\cite{maurizio2025quantumcomputinggenomicsconceptual} for a recent review), and the resolution of tangles in pangenome graphs which requires aligned genomic data as input. On this latter use case, the Quantum Approximate Optimisation Algorithm (QAOA)~\cite{farhi_quantum_2014}, promises to accelerate the pangenome graph tangle resolution and provide general solutions that are currently unobtainable classically~\cite{Cudby_2025wtk, Cudby_2025nbs}.

To realise this potential, a scalable and efficient encoding of genomic data for quantum systems is essential. Quantum hardware is limited at present by noise, meaning only relatively low depth quantum circuits can be executed. As such, simulation is a crucial step to demonstrate the scalability of any encoding on the gate-based quantum systems of the near future. With various proposed methods, some deterministic and utilising techniques such as the Cosine-Sine Decomposition \cite{niemann_logic_2016, shende_synthesis_2006}, and others probabilistic, employing techniques such as a genetic algorithm for state preparation (GASP) \cite{creevey_gasp_2023}. Maurizio, and Mazzola in Ref.~\cite{maurizio2025quantumcomputinggenomicsconceptual} cast doubt on the scalability of quantum bioinformatics applications due to data loading, but as found in Ref.~\cite{Creevey:2025qsr}, the loading of data at $100\%$ fidelity may not be necessary.

\begin{figure*}
  \includegraphics[width=0.95\linewidth]{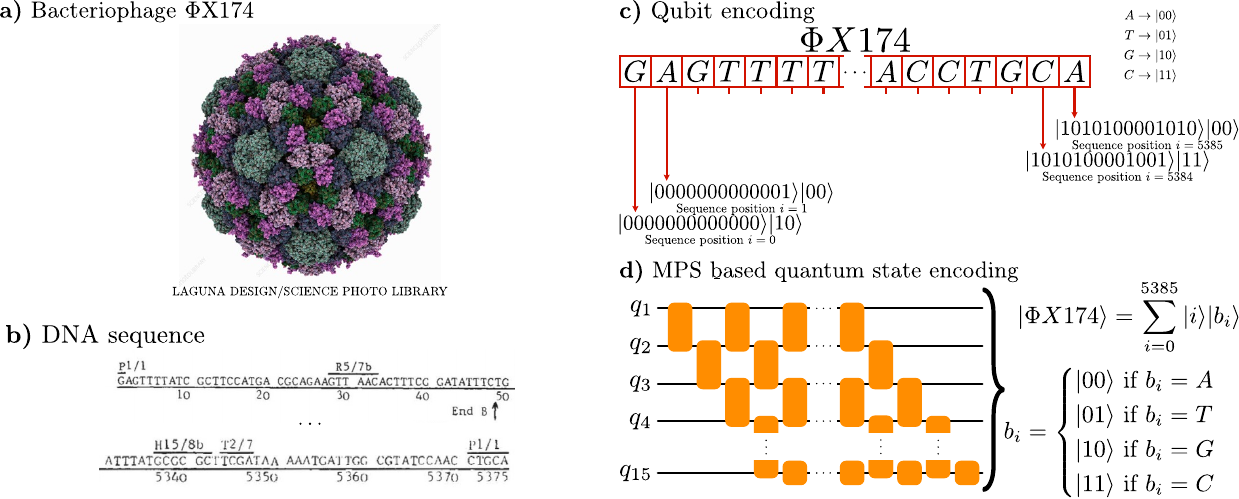}
  \caption{\textbf{a)} Illustration of the bacteriophage $\Phi X174$ from Ref.~\cite{Sanger_1977}. \textbf{b)} DNA sequence of $\Phi X174$ as listed in Ref.~\cite{Sanger_1977}. \textbf{c)} Procedure for encoding genomic data into a quantum state vector using the nucleotide base alphabet of Eq.~(\ref{eq:method_genome_repr}) (i.e.~a $k$-mer encoding of length $k=1$). For a larger value of $k$, the number of bases entering each component of the superposition will increase to $k$. \textbf{d)} Structure of circuits generated with the MPS method for $\Phi$X$174$, a $15$-site MPS (with qubits labelled $1$ to $15$). The structure is composed of layers of $14$ layered two-qubit unitaries connecting neighbouring qubits from qubit $1$ to qubit $15$, and a single-qubit unitary gate on qubit $15$.}
  \label{fig:intro_general_mps_circuit}
\end{figure*}

Tensor networks, specifically matrix product states (MPS), have recently gained popularity for simulating quantum circuits due to their accuracy and computational efficiency~\cite{ran_encoding_2020, rudolph_decomposition_2023, malz_preparation_2024, ben-dov_approximate_2024, smith_constant-depth_2024, green2025quantumencodingstructureddata}. However, most research is currently centred on simulating quantum circuits as MPS, rather than preparing MPS with quantum circuits. The method presented in this study produces quantum circuits that prepare states representing MPS to high fidelity, complementing the efficient state-vector encoding of genomic data that is used as the target MPS for the circuit. Importantly, the method itself is not uniquely applicable to encoded genomic data, but can prepare efficient circuits for arbitrary normalised quantum state vectors - including simple states such as Dicke, $|w\rangle$, or Gaussian states, or states encoding less structured data. The circuits produced scale favourably in both depth and total gate count. A high-level schematic of a circuit produced in this way is shown in Figure~\ref{fig:intro_general_mps_circuit}. As shown, this method produces circuits with sequential layers of two-qubit unitary gates that entangle neighbouring qubits (and a single-qubit gate on the last qubit). The form of each gate is derived from the MPS representation of the given state vector.

% \begin{figure}
%   \includegraphics[width=0.95\linewidth]{figures/circuit_ansatz.png}
%   \caption{Structure of circuits generated with the MPS method for an $n$-site MPS (with qubits labelled $1$ to $n$). The structure is composed of layers of $n-1$ layered two-qubit unitaries connecting neighbouring qubits from qubit $1$ to qubit $n$, and a single-qubit unitary gate on qubit $n$.}
%   \label{fig:intro_general_mps_circuit}
% \end{figure}

This method of state preparation is then applied to the use case of genomic data, resulting in the first quantum circuits that prepare states representing genome sequence reads on quantum devices. The scalability of this approach is demonstrated with simulated circuits on high performance computing (HPC) clusters, and with circuits executed on current quantum hardware.

%% file: sections/method.tex
\section{Methodology} \label{sec:method}

The method used to produce scalable quantum circuits to prepare states is applied to genomic data. To demonstrate this use case, it is necessary to outline how genomic data can be uniquely mapped to normalised quantum state vectors.

% \textcolor{red}{@Floyd - I've tried to flesh out the definitions a bit more following Sergii's comments, please could you check if what I've written is sensible (!):  added a definition of reconstruction error, mentioned how algorithmic fidelity ties into hardware fidelity in non-trivial ways, and discussed how the method can be adapted to other hardware given an understanding of how these two interact}

Throughout this section the following definitions are adopted:
\begin{enumerate}
    \item The \textit{target state} $\ket{\psi}$ is the state which is to be prepared by a quantum circuit.
    \item The \textit{prepared state} $\ket{\phi}$ is the approximation to the target state, produced by evolving the zero state $\ket{0}$ through the quantum circuit which is derived to prepare the target state.
    % \item The \textit{fidelity} between any two quantum states $\ket{\psi}$ and $\ket{\phi}$ is given by: $|\langle\psi|\phi\rangle|^2$.
    \item The \textit{reconstruction error}\footnote{also often referred to as the \textit{relative infidelity} in the literature.} between a state $\ket{\phi}$ prepared by a quantum circuit to approximate a target state $\ket{\psi}$ is given by: $1 - |\langle\psi|\phi\rangle|^2$ (i.e.~the difference from unity of the \textit{fidelity} between the two states, $|\langle\psi|\phi\rangle|^2$).
\end{enumerate}

\noindent The distinction is further drawn between two aspects of fidelity that become useful to decouple in the following discussion.

\begin{enumerate}
    \item The \textit{algorithmic fidelity} is the fidelity of the \textit{prepared state} to the \textit{target state} arising from the \textbf{algorithm} that constructs the circuit (i.e.~for a noiseless quantum device).
    \item The \textit{hardware fidelity} is the fidelity of the \textit{prepared state} to the \textit{target state} resulting from execution on \textbf{quantum hardware}.
\end{enumerate}

\noindent This gives rise to related notions of reconstruction error arising from algorithmic and hardware considerations. Additionally, when producing MPS representations of quantum states the bond dimension (see Sec.~\ref{sec:method_producing_mps} for details) can be reduced to produce a more compressed encoding, meaning that the MPS representation itself can have a reduced fidelity to the target state. This is briefly explored in Sec.~\ref{sec:results-hpc}, though we in general we use MPS with high enough bond dimension to ensure 100\% fidelity to the target state.

Since the circuit to produce the target state will prepare that state to a certain algorithmic fidelity in a noiseless environment, the hardware fidelity reported by its execution on noisy hardware is necessarily lower. The impact of noise on a circuit with given algorithmic fidelity is non-trivial (this is explored in Secs.~\ref{sec:results-hpc}-\ref{sec:results-hardware}). A significant advantage of the method presented in this section is that increasing the gate count of the circuits produced directly increases the algorithmic fidelity of the prepared state to the target state. This means that the algorithmic error is stable and highly controllable, allowing for adaptation of the approach to hardware platforms of differing noise levels. In practice, this means producing quantum circuits that prepare the target state to differing algorithmic fidelity, contingent upon an understanding of how the hardware fidelity varies as a function of the algorithmic fidelity (see Figure~\ref{fig:simulator_results} for such a breakdown for small test cases). 

\subsection{Encoding genomic data} \label{sec:method_encoding_data}

A genome sequence read is a string drawing from an alphabet of size four, where each letter represents a nucleotide base. Using a register to encode the position $i$ of the base within a genome read, and a register for the type of base $b_i$ drawing from an alphabet $\Sigma$, a genome read $g$ of length $L$ can be represented as
\begin{equation}
    g = b_0 b_1 \dots b_{L-2}b_{L-1} \qquad b_i \in \Sigma.
    \label{eq:method_genome_repr}
\end{equation}
For DNA, the alphabet $\Sigma = \{A, T, C, G\}$ is used. This formulation allows for a simple mapping to a quantum state representation by decomposing the genome read into two quantum registers~\cite{hollenberg_fast_2000} as,
\begin{equation}
    |\psi\rangle = \frac{1}{\sqrt{L}}\sum_{i=0}^{L-1}|i\rangle|b_i\rangle, \ b_i = \begin{cases}
        |00\rangle &\text{if $b_i=A$,}\\
        |01\rangle &\text{if $b_i=T$,}\\
        |10\rangle &\text{if $b_i=G$,}\\
        |11\rangle &\text{if $b_i=C$.}
    \end{cases}
    \label{eq:method_basic_encoding}
\end{equation}
For example, the short genome read
\begin{equation*}
    g = ATGC,
\end{equation*}
would be represented by the quantum state
\begin{align*}
    |\psi\rangle &= \frac{1}{\sqrt{4}}(|000\rangle|00\rangle + |001\rangle|01\rangle + |010\rangle|10\rangle + |011\rangle|11\rangle) \\
    &=: \frac{1}{\sqrt{4}}(|00000\rangle + |00101\rangle + |01010\rangle + |01111\rangle).
\end{align*}
Notably, the encoding displayed is part of a larger family of encodings, based not on individual bases but on $k$-mers i.e.~the unique genome sequence reads of length $k$. If instead of the alphabet of Eq.~(\ref{eq:method_genome_repr}), the alphabet was defined by the unique $k$-mers of a fixed length $k$ then the number of qubits required for the alphabet register would increase. The procedure with $k=1$ i.e.~with an alphabet drawn only from bases is detailed diagrammatically in Figure \ref{fig:intro_general_mps_circuit} \textbf{c)}.
% \begin{figure}
%   \includegraphics[width=0.95\linewidth]{figures/mps_encoding_fig.pdf}
%   \caption{Procedure for encoding genomic data into a quantum state vector using the nucleotide base alphabet of Eq.~(\ref{eq:method_genome_repr}) (i.e.~a $k$-mer encoding of length $k=1$). For a larger value of $k$, the number of bases entering each component of the superposition will increase to $k$.}
%   \label{fig:bio_encoding}
% \end{figure}
The number of qubits $n$ required to encode this data is the sum of those required for the position and base registers:
\begin{equation}
    n = \lceil\log_2(L)\rceil + \log_2(|b_i|).
    \label{eq:method_qubit_scaling}
\end{equation}
The utility of this encoding method is demonstrated with the genome of the bacteriophage $\Phi X174$ (the first ever fully sequenced genome, by Fred Sanger and collaborators in 1977 \cite{Sanger_1977}). The genome of $\Phi X174$ is short (with a length of 5386 base pairs) and is frequently used as a `standard genome' to benchmark next-generation sequencing (NGS) devices in genomics~\cite{Hardwick:2017ngs}. In the quantum encoding of Eq.~(\ref{eq:method_basic_encoding}), the genome of $\Phi X174$ can be encoded with 15 qubits. This makes the genome a suitable candidate for benchmarking quantum bioinformatics algorithms, such as QSA, to evaluate their use on real biological data. Similarly – highlighting a potent use case for modern pathogen surveillance with direct consequences to human health – the gene encoding the SARS-CoV2 spike protein (of length roughly 2000 base pairs~\cite{ncbi_sars_cov2_s_gene}) can be encoded in 14 qubits with the described method. This presents an `application-ready` demonstration of quantum bioinformatics algorithms that will highlight the viability of future methods to resolve biological problems.

\subsection{Representing quantum states as MPS}\label{sec:method_producing_mps}

To enable the construction of quantum circuits, the state vector encoding the genomic data is then transformed into a MPS, which are tensor networks whose structure and scalability suit them well to represent quantum states. The specific form of the MPS used is left-canonical MPS, i.e~one for which all $U$ matrices obtained from the decomposition of the state are left-orthonormal, satisfying,
\begin{equation*}
    \sum_{\sigma_i}(U^{\sigma_i})^\dag U^{\sigma_i} = \mathbb{I}.
\end{equation*}
The procedure for producing a left-canonical MPS is described in detail in Ref.~\cite{schollwoeck_density-matrix_2011}, and is outlined briefly here. An arbitrary quantum state can be converted to a left-canonical MPS via repeated applications of the matrix singular value decomposition. Starting from an arbitrary quantum state,

\begin{equation}
    |\psi\rangle = \sum_{\sigma_1,\ldots,\sigma_n}T^{\sigma_1\ldots\sigma_n}|\sigma_1, \ldots, \sigma_n\rangle
\end{equation}

the tensor $T^{\sigma_1\ldots\sigma_n}$ is reshaped as $T^{\sigma_1,(\sigma_2\ldots\sigma_n)}$, and a singular value decomposition is taken,

\begin{equation}
    \begin{aligned}
            T^{\sigma_1,(\sigma_2\ldots\sigma_n)} &= \sum_{a_1}^{r_1}U_{\sigma_1, a_1}S_{a_1, a_1}V^\dag_{a_1, (\sigma_2\ldots\sigma_n)}\\
            &\equiv\sum_{a_1}^{r_1}A^{\sigma_1}_{a_1}T^{(a_1, \sigma_2),(\sigma_3\ldots\sigma_n)},
    \end{aligned}
\end{equation}

Where $A^{\sigma_1}_{a_1} = U_{\sigma_1, a_1}$ and $T^{(a_1, \sigma_2),(\sigma_3\ldots\sigma_n)}=S_{a_1, a_1}V^\dag_{a_1, (\sigma_2\ldots\sigma_n)}$. Continuing this process with further SVDs and reshaping yields,

\begin{equation}
    T^{\sigma_1\ldots\sigma_n} =\sum_{a_1, \ldots, a_{n -1}}A^{\sigma_1}_{a_1}A^{\sigma_2}_{a_1, a_2}\ldots A^{\sigma_{n-1}}_{a_{n-2}, a_{n-1}}A^{\sigma_n}_{a_{n-1}},
\end{equation}

or more compactly,

\begin{equation}
    T^{\sigma_1\ldots\sigma_n} = A^{\sigma_1}A^{\sigma_2}\ldots A^{\sigma_{n-1}}A^{\sigma_n},
\end{equation}

Allowing,

\begin{equation}
    |\psi\rangle =A^{\sigma_1}A^{\sigma_2}\ldots A^{\sigma_{n-1}}A^{\sigma_n}|\sigma_1\ldots\sigma_n\rangle.
\end{equation}

\begin{figure}
  \includegraphics[width=0.95\linewidth]{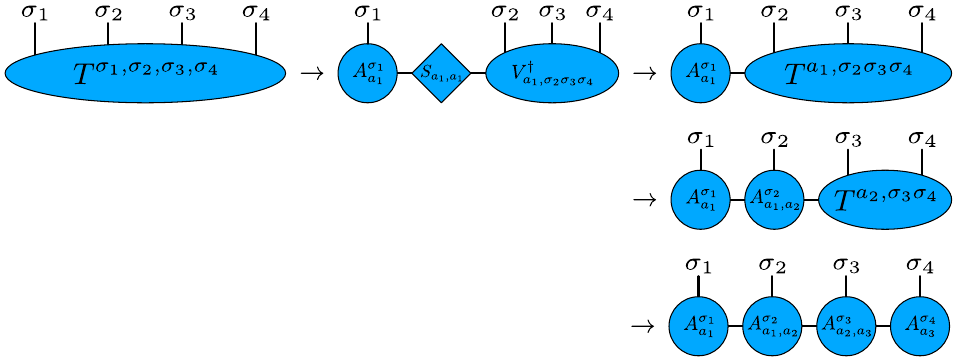}
  \caption{The procedure for converting a quantum state (whose tensor representation has rank four) to a matrix product state.}
  \label{fig:state_to_mps}
\end{figure}

This procedure can be represented compactly with Penrose notation for tensor operations, an illustration of this for a toy example depicting a rank-4 tensor is displayed in Figure~\ref{fig:state_to_mps}.

Rewriting $\ket{\psi}$ as a MPS with $n$ sites gives
\begin{equation}
    |\psi\rangle = A^{\sigma_1} A^{\sigma_2}\ldots A^{\sigma_{n-1}} A^{\sigma_n} \ket{\sigma_1, \ldots, \sigma_n},
    \label{eq:method_mps_repr}
\end{equation}
where each $A^{\sigma_i}$ (with $\sigma_i$ denoting the \textit{physical index}) is a complex matrix of shape $(\chi^i, \chi^{i+1})$. The internal dimensions $\chi^i$ describe the entanglement between neighbouring sites and are referred to as \textit{bond dimensions}. The maximum bond dimension $\chi$ of a MPS is often referred to just as the bond dimension and expresses the maximum entanglement.

Any quantum state can be represented as a MPS whose bond dimension scales as $2^{n/2}$ with the number of qubits $n$ required to encode the data (i.e.~an exponential scaling). Each \textit{site} in the MPS represents a qubit i.e.~the tensors $A^{\sigma_i}$ have physical dimension two. The bond dimension expresses the entanglement between qubits. The states considered in this study are highly entangled (as can be seen from the required bond dimension for reconstruction in Fig. \ref{fig:reconstruction_error}) and thus cannot be expressed as translationally-invariant MPS with constant bond dimension.

\subsection{Producing scalable quantum circuits}\label{sec:method_mps_to_circuit}

The encoding of genome sequence data provides one of the first use-cases of MPS state preparation techniques in genomics, demonstrating the efficacy of MPS state preparation for quantum systems. As mentioned in Sec.~\ref{sec:introduction}, this method produces circuits that evolve the zero state to a target MPS (i.e.~that which encodes the genomic data). The method demonstrated in this study is similar to and based on methods such as those of ref.~\cite{Iaconis:2023rna}, except that the inverse circuit is prepared -- the circuit which evolves the target state to the zero state. Solving the inverse problem gives significant computational efficiency advantages since the convergence of the state in each recursion in the algorithm only needs to be compared to the zero state. This saves the need to compare the prepared state at each recursion step against the target state which is computationally expensive, and allows for rapid generation and simulation of large qubit states (see Sec.~\ref{sec:results-hpc} for details).

\begin{figure}
  \includegraphics[width=0.95\linewidth]{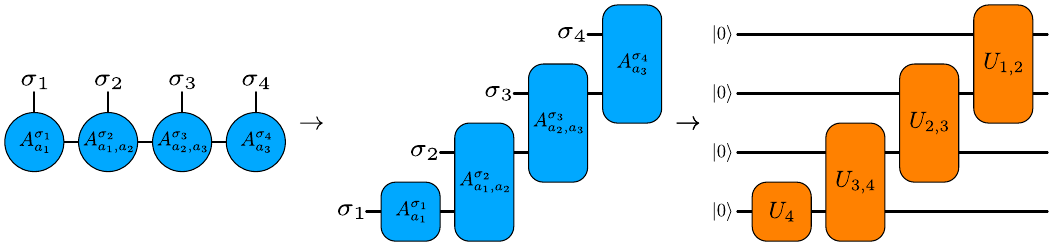}
  \caption{The circuit obtained for a truncated MPS representation, with $\chi=2$, showing a single layer of unitary gates for a toy four-qubit system. The gates are derived according to Eqs.~(\ref{eq:method_unitary_gates}) and~(\ref{eq:method_last_unitary}).}
  \label{fig:method_bond_dim_two_circuit}
\end{figure}

Firstly, the MPS must be converted to a quantum circuit with $\chi=2$. To achieve this, the rank of the MPS is increased to construct entangling unitary gates, starting from the $n^\mathrm{th}$ qubit. For the  $k^\mathrm{th}$ site these gates are denoted $U_{\sigma_k, \sigma_{k+1}}$, and are given by,\footnote{In Eq.~(\ref{eq:method_unitary_gates}), the null space of a (complex) square matrix $V$ (of dimension $d$) is defined as:
\begin{equation*}
    \operatorname{Null}\left \{V\right \} = \left\{ x \in \mathbb{C}^d : Vx = 0\right\}.
\end{equation*}}
\begin{align}
    V_{\sigma_k, a_{k-1}} &= A^{\sigma_k}_{a_k-1}, \\
    U_{\sigma_k, \sigma_{k-1}} &= \left [ V_{\sigma_k, \sigma_{k-1}}, \operatorname{Null}\left \{V_{\sigma_k, \sigma_{k-1}}\right \}\right ].
    \label{eq:method_unitary_gates}
\end{align}
The formulation of Eq.~(\ref{eq:method_unitary_gates}) applies to the gates $U_{1,2}$ to $U_{n-1, n}$, though the single-qubit gate on the $n^\mathrm{th}$ qubit is given by
\begin{equation}
    U_n = A^{\sigma_1}_{a_1}.
    \label{eq:method_last_unitary}
\end{equation}
This procedure can be demonstrated compactly in Penrose notation~\cite{Penrose:1971ndt} as shown in Figure~\ref{fig:method_bond_dim_two_circuit}.
To generalise for target MPS $\ket{\psi}$ of arbitrary bond dimension $\chi > 2$, this simpler method is employed in an iterative process which adds layers of unitary operators to the circuit $\mathcal{C}$. This circuit is used to evolve the zero state to the target MPS, and successive iterations improve the fidelity between the prepared and target states.

This iterative procedure is outlined below. Assuming there is a required fidelity threshold $f$ between the target and evolved states, the algorithm follows,
% Let $\ket{\psi_q}$ be the evolved state at iteration $q$ of the general algorithm (defining $\ket{\psi_0}\equiv\ket{\psi}$), then:

\begin{tcolorbox}
    [title={MPS State Preparation}]
    \emph{Input:} the prepared state at iteration $q$ of the general algorithm (defining $\ket{\psi_0}\equiv\ket{\psi}$), $\ket{\psi_q}$, and the fidelity threshold $f$ for the prepared state to satisfy.\\
    \emph{Output:} the quantum circuit $\mathcal{C}^q$.
    \begin{enumerate}
        \itemsep-0.2em 
        \item Compute $\ket{\tilde{\psi}_q}$ by truncating the MPS representation of $\ket{\psi_q}$ to bond dimension two, and using the procedure outlined in Eqs.~(\ref{eq:method_unitary_gates}) and~(\ref{eq:method_last_unitary}) to calculate the gates $U_{\sigma_k, \sigma_{k-1}}$. This creates the circuit $\mathcal{C}^q$ which evolves the zero state to $\tilde{\ket{\psi_q}}$ as $\mathcal{C}^q \ket{0} = \tilde{\ket{\psi_q}}$.
        \item Use this circuit to produce the inverse circuit, $C^{q\dag}$, which would evolve $\tilde{\ket{\psi_q}}$ to the zero state as $\mathcal{C}^{q\dagger} \tilde{\ket{\psi_q}} = \ket{0}$.
        \item Let $\ket{\psi_{q+1}} = \mathcal{C}^{q\dagger} \ket{\psi_q}$. If the fidelity between the state $\ket{\psi_{q+1}}$ and the zero state is larger than or equal to $f$ then the circuit has been found, return $\mathcal{C}^q$.
        \item Otherwise, return to step 1.
    \end{enumerate}
\end{tcolorbox}

\begin{figure}
  \includegraphics[width=0.95\linewidth]{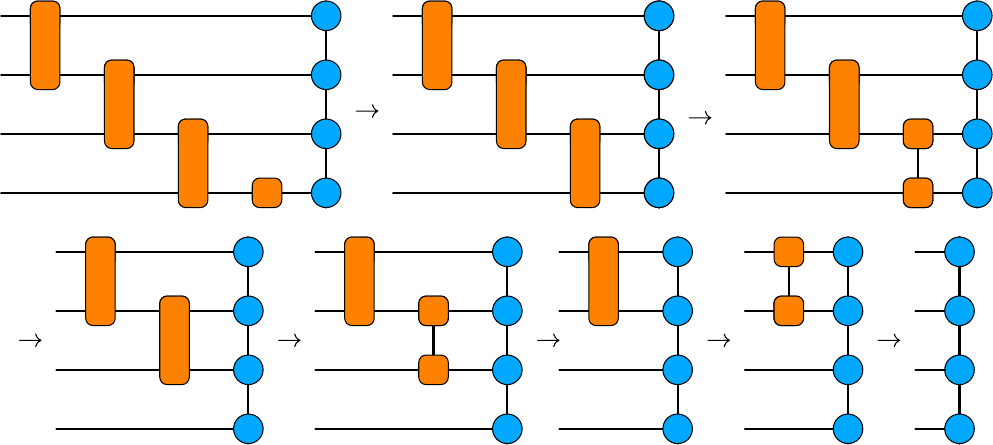}
  \caption{The contraction of gates into the MPS nodes required for each iteration of the algorithm requiring a bond dimension larger than 2.}
  \label{fig:mps_contraction}
\end{figure}

% \begin{enumerate}
%     \item Compute $\tilde{\ket{\psi_q}}$ by truncating the MPS representation of $\ket{\psi_q}$ to bond dimension two, and using the procedure outlined in Eqs.~(\ref{eq:method_unitary_gates}) and~(\ref{eq:method_last_unitary}) to calculate the gates $U_{\sigma_k, \sigma_{k-1}}$.
%     \item This creates the circuit $\mathcal{C}^q$ which evolves the zero state to $\tilde{\ket{\psi_q}}$ as $\mathcal{C}^q \ket{0} = \tilde{\ket{\psi_q}}$.
%     \item Use this circuit to produce the inverse circuit, evolving $\tilde{\ket{\psi_q}}$ to the zero state as $\mathcal{C}^{q\dagger} \tilde{\ket{\psi_q}} = \ket{0}$.
%     \item Let $\ket{\psi_{q+1}} = \mathcal{C}^{q\dagger} \tilde{\ket{\psi_q}}$. If the fidelity between the state $\ket{\psi_{q+1}}$ and the zero state is larger than or equal to $f$ then the circuit has been found, return $\mathcal{C}^q$.
%     \item Otherwise, return to step 1.
% \end{enumerate}
The process of generating the circuit begins from the target state, iteratively adding circuit layers which transform the given state back to the zero state. The inverse of this circuit evolves the zero state to the target state, this circuit will have $q$ layers and initialise $\ket{\psi}$ to an accuracy determined by the constraints imposed on the reconstruction error. This procedure is shown in Figure~\ref{fig:mps_contraction}, using Penrose notation to illustrate the absorption of gates into the MPS nodes for an iteration of the above algorithm.

It is emphasised again that this method of state preparation produces scalable quantum circuits for an arbitrary normalised target state. Thus circuits can be produced for any target state, and scale efficiently when simulated on HPC and executed on quantum hardware.

%% file: sections/complexity.tex
\section{Complexity Analysis}\label{sec:complexity}

The complexity of the method will be discussed in terms of total gate count, where $n$ is the total number of qubits, $\chi_{\rm{max}}$ is the maximum bond dimension of the produced MPS, and the reconstruction error is given by $\delta^2 = 1 - |\langle\psi|\phi\rangle|^2$.

The results of Ref.~\cite{Jobst:2023wdt} demonstrate a theoretical bound for the gate scaling of MPS representations of quantum circuits, with gate count $n_g$ scaling as,

\begin{equation}
    n_g \sim n \left [ \log_2\left ( \frac{1}{\epsilon}\right) \right ]^2,
    \label{eq:method_gate-count-scaling}
\end{equation}

\noindent where $\epsilon$ is given by the $l_2$ distance between $|\psi\rangle$ and $|\phi\rangle$, i.e. $\epsilon = |||\psi\rangle - |\phi\rangle||_2$. This relation is derived from considerations of how $\chi_{\rm{max}}$ scales with $\epsilon$. Expanding the square of this quantity,

\begin{align}
    |||\psi\rangle - |\phi\rangle||^2_2 &= (\langle\psi| - \langle\phi|)(|\psi\rangle -|\phi\rangle) \\ \nonumber
    &= 2(1- \rm{Re}\langle\psi|\phi\rangle),
    \label{eq:method_reconstruction-error}
\end{align}

\noindent the dependence on the overlap $\langle\psi|\phi\rangle$ becomes clear. This overlap can be made purely real by allowing a change in the global phase of $|\phi\rangle$. This demonstrates the relation between $\epsilon$ and $\delta$ directly,

\begin{align}
    |||\psi\rangle - |\phi\rangle||_2 \equiv \epsilon &= \sqrt{2(1 - \sqrt{1 - \delta^2})}.
    % \label{eq:method_epsilon-delta-relation}
\end{align}

\noindent In the limit of small reconstruction error, the inner square root scales as $\sqrt{1-\delta^2} \sim 1 - \delta^2/2 + \mathcal{O}(\delta^4)$, which gives the asymptotic scaling $\epsilon \sim \delta$.

The gate complexity of the circuits presented in the following sections scale (pre-transpilation) according to,

\begin{equation}
    n_g \sim n\cdot \chi_{\rm{max}}^2,
    \label{eq:method_gate-count-scaling-bond-dim}
\end{equation}

\noindent which scales consistently with the relation of Eq.~(\ref{eq:method_gate-count-scaling}) in terms of $\epsilon$, given the asymptotic relation of $\epsilon$ and $\delta$. Notably, Figure~\ref{fig:reconstruction_error} demonstrates explicitly the relation between the (maximum) bond dimension and $\delta^2$.

%% file: sections/results-hpc.tex
\section{Circuit Generation and Simulation on HPC} \label{sec:results-hpc}

The performance of the algorithm for constructing circuits to prepare MPS is evaluated by generating and simulating circuits on HPC. The calculations and results shown in this section were performed on a dedicated pool of HPC resources, comprising a four-way and an eight-way Nvidia H100 GPU host totaling $320$ CPU and 2TB in memory.

\begin{figure}[!htbp]
  \includegraphics[width=0.95\linewidth]{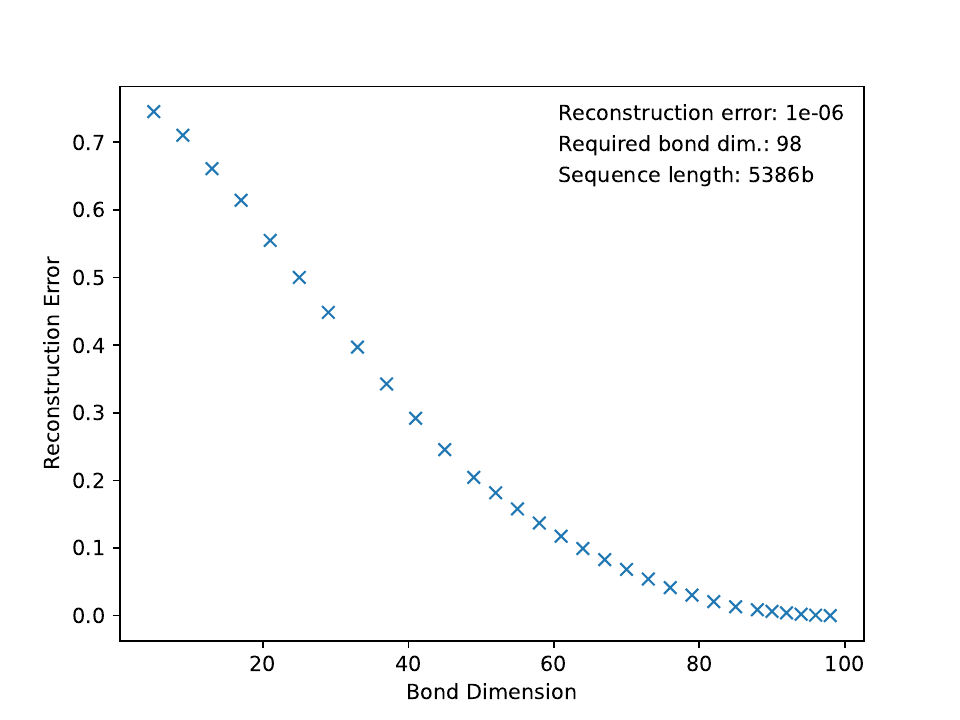}
  \caption{The reconstruction error for encoding the genome of the bacteriophage as a MPS against the maximum bond dimension of the MPS.}
  \label{fig:reconstruction_error}
\end{figure}

The first case explored is the scaling of the reconstruction error of the MPS encoding relative to the original state against the bond dimension of the MPS. The genome of the bacteriophage, $\Phi X174$ ($L=5386$), is used as a test case here, and it is again noted that this genome can be fully encoded in 15 qubits. The MPS that represents this genome has a bond dimension of $98$ to give a reconstruction error of $0.00001\%$, as shown in Figure~\ref{fig:reconstruction_error}. Even halving the bond dimension here gives a reconstruction error of approximately $20\%$.

It is important to emphasise that the reconstruction error discussed in Figure~\ref{fig:reconstruction_error} is that between the \textbf{state vector} and the \textbf{MPS representation} of a given bond dimension, indeed, the bond dimension required for zero reconstruction error of an arbitrary quantum state scales as $2^{n/2}$ for $n$ sites, in general.

\begin{figure}[!htbp]
  \includegraphics[width=0.95\linewidth]{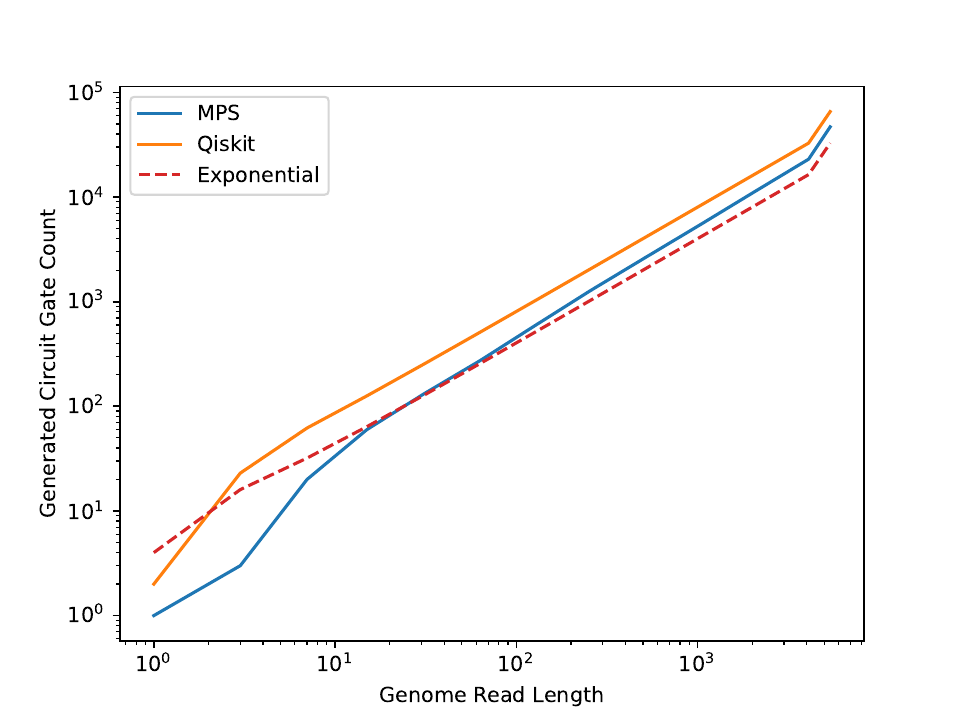}
  \caption{Gate scaling of the MPS technique compared to the initialisation functionality of IBM Qiskit, assessed on reads drawn from the $\Phi X174$ genome. The blue and orange lines represent the gate count of the MPS technique and IBM Qiskit respectively. The red dotted line shows a linear extrapolation from $2^n$ where $n$ is the number of qubits to encode the given genome read.}
  \label{fig:phi_scaling}
\end{figure}

The MPS technique produces circuits that scale better than using IBM Qiskit~\cite{Javadi-Abhari:2024kbf} (v1.1.1) to produce circuits of equivalent algorithmic fidelities, as shown for genome reads up to the length of $\Phi X174$ in Figure~\ref{fig:phi_scaling}. The basis used for the Qiskit decomposition was the set $\{u_3(\theta,\varphi,\psi), C_X\}$, which form a universal gate set (with functional forms as given in Ref.~\cite{Doi:2020cb}). The circuits produced with the MPS method were prepared to 99\% algorithmic fidelity.

\begin{figure}[!htbp]
  \includegraphics[width=0.95\linewidth]{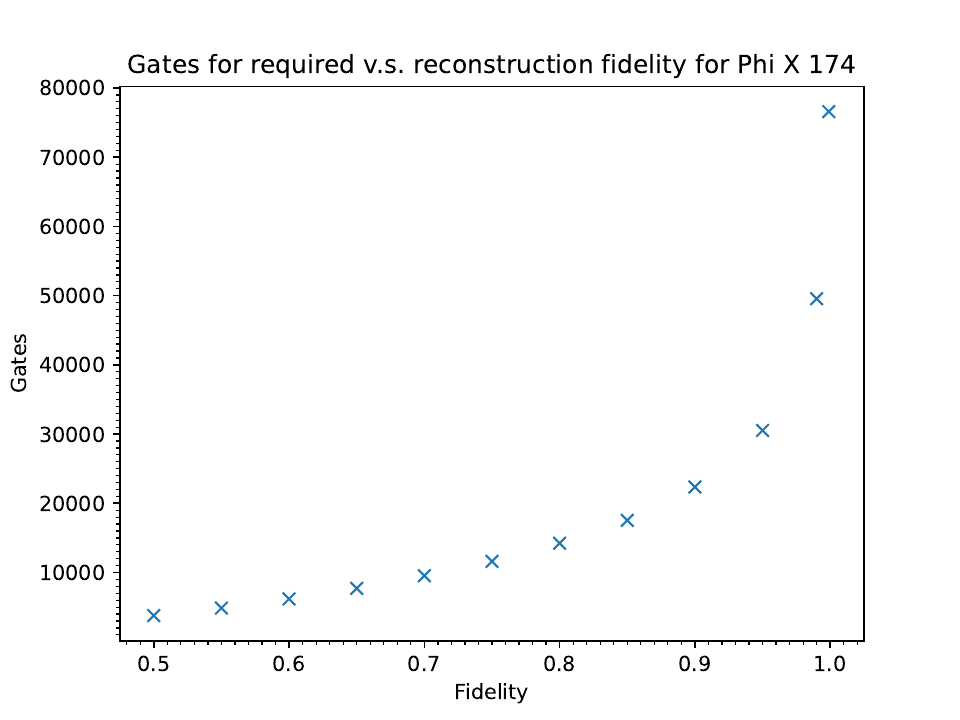}
  \caption{The gate count for the circuits encoding the bacteriophage $\Phi X174$ to specific algorithmic fidelities (between prepared and target states).}
  \label{fig:phix_gate_scaling}
\end{figure}

The scaling obtained is further illustrated in Figure~\ref{fig:phix_gate_scaling}, where the gate count required for the encoding of $\Phi X174$ is shown for a range of algorithmic fidelities. This could be further improved using depth-width tradeoffs, as well as using adaptive quantum circuits~\cite{smith_constant-depth_2024}, though this remains an active field of research and will be the focus of future studies.

For a circuit with $n$ qubits reconstructing the target MPS to a given algorithmic reconstruction error, the number of gates required scales approximately as given in Eq.~(\ref{eq:method_gate-count-scaling}), with fewer gates required as the algorithmic fidelity requirement is relaxed. It is important to note that though the MPS method scales approximately exponentially in gate count, $2^n$, it is encoded on a logarithmic number of qubits, $n = \lceil\log_2L\rceil + 2$, meaning the number of operations (gates) scales as a linear polynomial in genome read length.

\begin{figure}
  \includegraphics[width=0.95\linewidth]{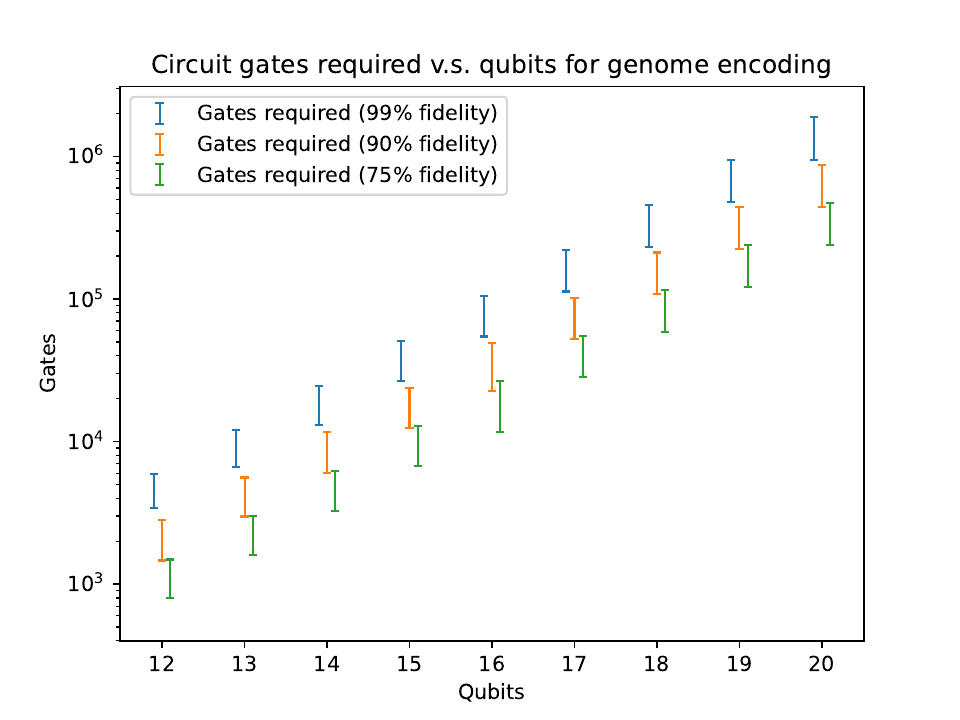}
  \caption{The ranges of gate count for circuits encoding genome reads of varying sizes to specific algorithmic fidelities (between prepared and target states), distributed against the number of qubits required for the encoding. The range of genome read lengths was chosen to be between the minimum and maximum to be encoded in the noted number of qubits (e.g.~between $513$ and $1023$ base pairs for the $12$-qubit circuit).}
  \label{fig:circuit_gate_count}
\end{figure}

A wide range of sequences of varying lengths have been simulated, with the results displayed in Figure~\ref{fig:circuit_gate_count}. This figure shows the number of gates required for circuits of $12-20$ qubits, based on the minimum length of genome read that can be encoded. The number of gates per qubit demonstrated follows a stable scaling of approximately $2^n$.

The results in Figure~\ref{fig:circuit_gate_count} demonstrate the scaling for encoding genomic data to various levels of fidelity.  Given the expected technology capabilities of current quantum hardware by the end of $2025$, it should be possible to run circuits which perform quantum sequence alignment (QSA) on quantum hardware for the S-Gene of SARS-CoV2, encoding the spike protein (requiring $14$ qubits), as well as analogous genes for similar pathogens affecting human health.
%\textcolor{red}{Furthermore, the `target region' of resource requirements specified in the Wellcome Leap Q4Bio project brief ($100-200$ qubits, circuit depth ~1M) would allow us to perform MSA for viral genomes (and genes of similar length $O(10kbp)$), for example, the full genome of SARS-CoV2  prepared to $99\%$ fidelity.} 
Notably, these expected increases in hardware capability should allow the entire genome of $\Phi X174$ to be encoded and sequenced on a physical quantum computer. This is a crucial milestone in the quantum analysis of pangenomes for individual genes and genomes. We highlight rough gate counts required to encode specific genes and genomes in Table~\ref{tab:mps_algorithmic_gates}.

\begin{table}
    \small\caption{Qubit and gate counts for states prepared for sequences at $75\%$ algorithmic fidelity. Rough values are shown in cases of high variability (notably SARS-CoV2 and HLA-DRB1)}
    \begin{tabular}{|c|c|c|c|}
     \hline
         \textbf{Sequence} & \textbf{Length} & \textbf{Qubits} & \textbf{Gate count (75\%} \\
          & & & \textbf{algorithmic fidelity)} \\
         \hline
         % \hline
         SARS-CoV2 S~\cite{ncbi_sars_cov2_s_gene} & ~1.2-2.1k & 13-14 & ~1.5k-5.1k \\
         \hline
         $\Phi X174$~\cite{Sanger_1977} & 5.386k & 15 & 11.610k \\
         \hline
         HLA-DRB1~\cite{ncbi_hla_drb1} & ~10-15k & 16 & ~11k-21k \\
         \hline
         SARS-CoV2~\cite{ncbi_sars_cov2} & ~30k & 17 & ~40k \\
         \hline
    \end{tabular}
    \label{tab:mps_algorithmic_gates} 
\end{table}

Large genomes, such as the human genome of length 3.2bn base pairs, can be analysed in multiple ways. Many applications require the full genome to be loaded at once (such as identifying large-scale structural variation), but others target specific regions of the genome (e.g.~analyses of individual genes). Pangenome graph tangle resolution and the phylogeny of specific genes fall into the latter category, e.g.~one may consider constructing the pangenome of the HLA-DRB1 gene to analyse variability in that region, rather than for the full human genome. Long-read sequencing results are themselves between 10,000 and 100,000 base pairs long, which the MPS state preparation method can fully encode, requiring 16-19 qubits. This qubit range is enough to encode fully most viruses including Influenza A~\cite{ncbi_influenza_a} and SARS-CoV2, as well as gram-negative plasmids from bacteria (understanding the pangenomic and phylogenetic evolution of these gives a doorway to tackling antibiotic resistance in harmful bacteria).

In the case where encoding large genomes fully, e.g.~of lengths L=250M+ base pairs (and thus 30+ qubits), is desired, justified approximations to reduce the circuit construction time and the required gate count can be made. Assuming that interactions within long genomes are confined to short ranges (where short is relative to the full length of the genome) of length $l=~70,000$ base pairs, the problem complexity can be effectively reduced and a convergent result can be obtained by translating this to a reduced long-range entanglement in the MPS. The effect of this is significant and reduces our gate count by a factor of roughly $L/l$ relative to the theoretical limit of $2^n$ for the $n$ qubits required to encode $L$ bases. This allows the simulation of circuits for large genomes in a feasible time on high performance computing clusters.

The assumptions made in simplifying the discussion for large genomes are well-justified when considering that the median gene length in the human genome is $~29,000$ base pairs, and that variations in large genomes across isolates occur near each other rather than separated by long distances. Quantum pangenomics is concerned with mapping variations across isolates and exploring this behaviour in phylogenetic tree construction and pangenome graph tangle resolution.
%However this method, while a strong proof-of-concept, is still in its infancy and would benefit from more theoretical development to fully leverage the capabilities of the MPS simulation suite.

Additionally, options for adaptive circuit generation (based on the work outlined in \cite{smith_constant-depth_2024}) for preparing quantum states have been explored. Applying such methods to translationally-variant MPS remains an open problem in quantum state preparation. Initial explorations of this field prove promising and could significantly decrease the gate-count requirements for preparing MPS for genomic data encoding. This would allow for circuits of constant depth to prepare the relevant states.

%% file: sections/results-hardware.tex
\section{Transpilation and Execution on Quantum Hardware Emulators}\label{sec:results-hardware}

Executing the circuits on quantum devices requires transpilation to the specific hardware. A benefit of the MPS encoding is that it is linear, meaning that it can be transpiled to any hardware efficiently, as it only requires entangling between neighbouring qubits. As a result of this, transpilation only involves converting the existing circuit into the native gate set of the given hardware, and does not have to rearrange qubits for the specific hardware connectivity. For example, given a native gate set,

\begin{equation}
    S = \{R_y(\theta), R_z(\theta), C_X\}
    \label{eq:transpilation_basis}
\end{equation}
Where,
\begin{equation*}
    R_z(\theta) = \begin{pmatrix}
        e^{-i\frac{\theta}{2}} & 0 \\
        0 & ae^{i\frac{\theta}{2}}
    \end{pmatrix}, R_y(\theta) = \begin{pmatrix}
        \cos\frac{\theta}{2} & -\sin\frac{\theta}{2} \\
        \sin\frac{\theta}{2} & \cos\frac{\theta}{2}
    \end{pmatrix},
\end{equation*}
\begin{equation*}
    C_X = \begin{pmatrix}
        1 & 0 & 0 & 0 \\
        0 & 1 & 0 & 0 \\
        0 & 0 & 0 & 1 \\
        0 & 0 & 1 & 0
    \end{pmatrix}.
\end{equation*}

An example of an efficient decomposition for an arbitrary two-qubit unitary gate using the basis of Eq.~(\ref{eq:transpilation_basis}) is shown in Figure~\ref{fig:gate_decomposition}.

\begin{figure*}[!htbp]
  \begin{quantikz}[thin lines,transparent,rounded corners,line width=0.25mm]
    & \gate[2]{U^{(2)}} & \\
    &&
  \end{quantikz}
    =
  \begin{quantikz}[thin lines,transparent,rounded corners,line width=0.25mm]
    & \gate{U^{(1)}} & \targ{1} & \gate{R_z(\delta)} & \ctrl[closed]{1} & \qw & \targ{1}  & \gate{U^{(1)}} & \\
    & \gate{U^{(1)}} & \ctrl[closed]{-1} & \gate{R_y(\beta)} & \targ{-1} & \gate{R_y(\alpha)} & \ctrl[closed]{-1} & \gate{U^{(1)}} &
  \end{quantikz}
  \par
  \vspace{1em}
  \begin{quantikz}[thin lines,transparent,rounded corners,line width=0.25mm]
    & \gate[1]{U^{(1)}} &
  \end{quantikz}
    =
  \begin{quantikz}[thin lines,transparent,rounded corners,line width=0.25mm]
    & \gate{R_z(\theta_0)} & \gate{R_y(\theta_1)} & \gate{R_z(\theta_2)} &
  \end{quantikz}
  \caption{Efficient $2$-qubit unitary decomposition into one- and two-qubit   gates in the gate set $\{C_X, R_y, R_z\}$.}
  \label{fig:gate_decomposition}
\end{figure*}

% Quantinuum hardware has a specific gate in their native gate set that allows for an efficient transpilation, this is referred to as the $TK2$ gate~\cite{quantinuum:2025pyt}

% \begin{equation*}
%     TK2(\alpha, \beta, \gamma)=e^{-\frac{1}{2}i\pi(\alpha(\hat{X}\otimes\hat{X})-\beta(\hat{Y}\otimes\hat{Y})-\gamma(\hat{Z}\otimes\hat{Z})}
% \end{equation*}

% Using this gate, transpilation to Quantinuum’s hardware alters the number of gates required for each circuit, as shown in Table \ref{tab:mps_hardware_gates}.

% \begin{table}
%     \small\caption{Pre- and post-transpilation gate counts for $\Phi X174$ at various algorithmic fidelities}
%     \begin{tabular}{|c|c|c|}
%      \hline
%      \textbf{Algorithmic} & \textbf{Pre-transpilation} & \textbf{Post-transpilation} \\
%      \textbf{Fidelity} & & \\
%      \hline
%      % \hline
%      70\% & 9706 & 27163 \\
%      \hline
%      80\% & 14641 & 40981 \\
%      \hline
%      90\% & 22921 & 64165 \\
%      \hline
%      99\% & 50446 & 141235 \\
%      \hline
%     \end{tabular}
%     \label{tab:mps_hardware_gates}
% \end{table}

\begin{figure*}[!htbp]
  \includegraphics[width=0.95\linewidth]{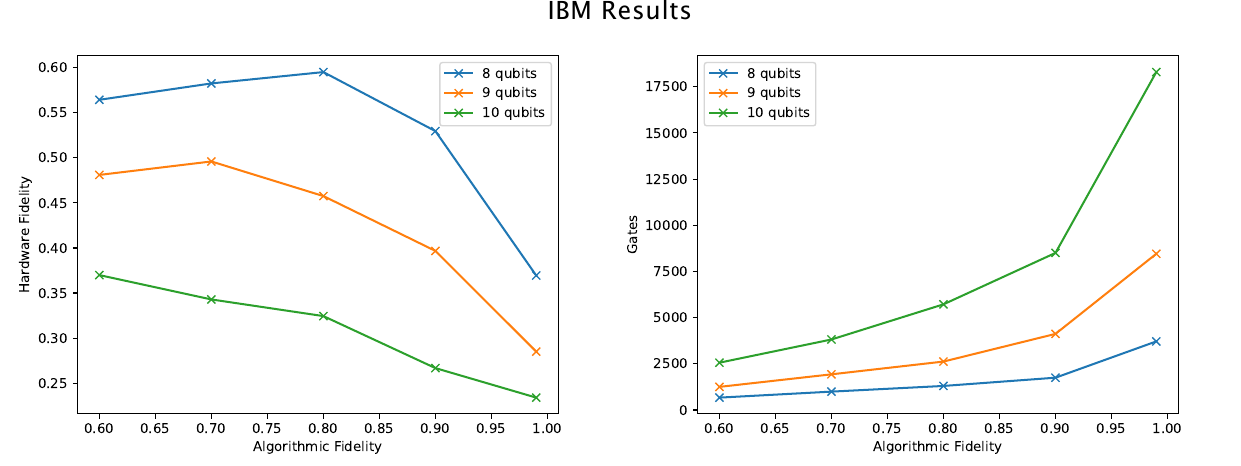}
  \caption{Results for $8$, $9$, and $10$ qubit tests on the IBM emulators. The first column displays the fidelity results, hardware fidelity being $|\langle\psi|\phi\rangle|^2$ where $|\psi\rangle$ is the state vector prepared to a given algorithmic fidelity, and $|\phi\rangle$ is the quasi-state vector constructed from the resultant shots distribution. The second column displays information about the circuits in terms of gates and depth.}
  \label{fig:simulator_results}
\end{figure*}

Initial small circuit tests of $8$, $9$, and $10$ qubits were executed on IBM hardware emulators, with results displayed in Figure~\ref{fig:simulator_results}. The results reveal a complex interplay between algorithmic fidelity and hardware fidelity. In general, improving the algorithmic fidelity requires circuits with higher gate counts, meaning that the corresponding increase in hardware fidelity continues up to a limit. This limit is the circuit gate count that can be supported by the hardware; if gates are added close to this limit, then the increase in noise causes the hardware fidelity to decrease.

 %This trend emphasises the importance of considering hardware roadmaps such as the IBM hardware roadmap shown in Figure~\ref{fig:IBM_roadmap} in the Appendix.
%Figure~\ref{fig:quantinuum_roadmap}

% From the trends shown in the Figure, it is evident that transpiled circuits for the cases outlined in Table~\ref{tab:mps_hardware_gates} are within reach in the near future for a NISQ computer.

%% file: sections/conclusions.tex
\section{Conclusions and Further Work}\label{sec:conclusions}

The method described in Section~\ref{sec:method} for constructing circuits to prepare MPS has been demonstrated to be efficient and scalable - surpassing the capabilities of standard tools for quantum state preparation. The approximately exponential gate count scaling of the MPS technique implies that the $\Phi X174$ genome has close to no redundancy, i.e.~it is already close to optimally compressed in representation. This further supports the efficiency of the MPS method, and supports the theoretical complexity analysis of the circuits detailed in Sec.~\ref{sec:complexity}.

Generating and simulating circuits with this method on HPC has illustrated the utility of this procedure. Exploring the usage of the method on quantum hardware in Section~\ref{sec:results-hardware} has shown that the prospect of encoding and working with genomic data on quantum systems is realistic in the near future. The use case of genomic data has been highlighted to demonstrate that significant problems of biological importance can be addressed on quantum hardware in the near future.

Additional work is planned to further develop the algorithm. One of these improvements is to examine the use of adaptive circuit generation (based on the work outlined in \cite{smith_constant-depth_2024}) for preparing quantum states to reduce the gate count. Applying such methods to translationally-variant MPS remains an open problem in quantum state preparation. Initial explorations of this field prove promising and could significantly decrease the gate-count requirements for preparing MPS for genomic data encoding, and in general. This would allow for circuits of constant depth to prepare the relevant states.

In addition to algorithmic developments, the application of the state preparation to algorithms for bioinformatics applications - such as sequence alignment - is in progress. An algorithm for quantum sequence alignment based on the formalism of ref.~\cite{hollenberg_fast_2000}, and recently implemented in~\cite{Creevey:2025qsr} is in development, leveraging the MPS state preparation method outlined for genomic data. Bolstering this effort will be implementing the state preparation method as part of quantum machine-learning (QML) approaches for encoding data, and for integrating with the work of \cite{creevey2023kernelalignmentquantumsupport,Yakymenko_2026iul}. This will lay the groundwork for fully quantum bioinformatics pipelines in the future - since genomics applications such as phylogenetic tree construction and pangenome graph tangle resolution depend upon prepared and aligned genomic data.

%% file: sections/acknowledgements.tex
\section{Acknowledgments}\label{sec:acknowledgements}

The authors would like to thank the rest of the QPG collaboration for many insightful discussions. The authors also express special thanks to Thorsten Wahl for helpful discussions and expertise with tensor networks. This work is supported by Wellcome Leap as part of the Q4Bio Program. This research was supported by the University of Melbourne through the establishment of the IBM Quantum Network Hub and the Australian Research Council Centre of Excellence for Quantum Biotechnology (CE230100021) at the University. SS acknowledges support from the Royal Society University Research Fellowship.
%The authors were also funded in part by the Wellcome Trust [Grant number 220540/Z/20/A].

% Sergii suggested to cut this part out, I guess it depends on the journal we are targetting - Hitham
\section{Author contributions statement}
   F.M.C. conceived the project with input from S.S. The computational framework was created by F.M.C. and H.T.H. who jointly undertook the computational work to produce the results presented in this paper, with input from all authors. All authors had input in writing the manuscript.

\section{Competing interests} \label{sec:interests}
    The authors declare no competing interests.

\section{Data availability} \label{sec:data}
    The datasets generated during and/or analysed during the current study are available from the corresponding author on reasonable request.